\documentclass[preprint2]{aastex}

\newcommand{\OII}{\mbox{[\ion{O}{2}]$\lambda\lambda 3726, 3729$\,\AA}}
\newcommand{\NII}{\mbox{[\ion{N}{2}]$\lambda6584$}\, \AA}

\newcommand{\ha}{\relax \ifmmode {\mbox H}\alpha\else H$\alpha$\fi}
\newcommand{\hb}{\relax \ifmmode {\mbox H}\beta\else H$\beta$\fi}
\newcommand{\sii}{\relax \ifmmode {\mbox S\,{\scshape ii}}\else S\,{\scshape ii}\fi}
\newcommand{\nii}{\relax \ifmmode {\mbox N\,{\scshape ii}}\else N\,{\scshape ii}\fi}
\newcommand{\oii}{\relax \ifmmode {\mbox O\,{\scshape ii}}\else O\,{\scshape ii}\fi}
\newcommand{\oiii}{\relax \ifmmode {\mbox O\,{\scshape iii}}\else O\,{\scshape 
iii}\fi}

\slugcomment{Submitted for publication in {\it The Astrophysical Journal Letters}}

\shortauthors{Amor\'in et al.}

\begin{document}

\title{On the oxygen and nitrogen chemical abundances and the evolution of the ``green pea'' galaxies}

\author{Ricardo O. Amor\'in}
\email{amorin@iaa.es}
\author{Enrique P\'erez-Montero}
\email{epm@iaa.es}

\and

\author{J. M. V\'ilchez}
\email{jvm@iaa.es}
\affil{Instituto de Astrof\'isica de Andaluc\'ia (CSIC), Glorieta 
de la Astronom\'ia S/N, E-18008 Granada,
    Spain}

\begin{abstract}
We have investigated the oxygen and nitrogen chemical abundances in
extremely compact star-forming galaxies (SFGs) with redshifts between
$\sim$0.11 and 0.35, popularly referred to as ``green peas''.  Direct and
strong-line methods sensitive to the N/O ratio applied to their Sloan 
Digital Sky Survey (SDSS) spectra reveal that these systems are genuine 
metal-poor galaxies, with mean oxygen abundances $\sim 20$\% solar.  
At a given metallicity
these galaxies display systematically large N/O ratios compared to
normal galaxies, which can explain the strong difference between our
metallicities measurements and previous ones.  While their N/O ratios
follow the relation with stellar mass of local SFGs in the SDSS, we 
find that the mass--metallicity relation of the
``green peas'' is offset $\ga$0.3 dex to lower metallicities.  We
argue that recent interaction-induced inflow of gas, possibly coupled
with a selective metal-rich gas loss, driven by supernova winds, may
explain our findings and the known galaxy properties, namely high
specific star formation rates, extreme compactness, and disturbed
optical morphologies.  The ``green pea'' galaxy properties seem to be
not common in the nearby universe, suggesting a short and extreme
stage of their evolution. Therefore, these galaxies may allow us to
study in great detail many processes, such as starburst activity and
chemical enrichment, under physical conditions approaching those in
galaxies at higher redshifts.

\end{abstract}


\section{INTRODUCTION}

The cosmological relevance of local starbursts is due to their
similarities with their analogs at high redshift. They may therefore
provide extremely valuable laboratories to study with high resolution
and sensitivity relevant physical processes associated with the
starburst activity in much better detail.  One of the most suitable
environments for these purposes can be the intriguing class of very
compact, extremely star-forming galaxy (SFG) at low redshift ($0.11 \la z
\la 0.35$) recently discovered by volunteers in the ``Galaxy Zoo''
project \citep{Lintott}.  These galaxies, popularly referred to as
``green peas'' (hereafter GPs), were first reported and studied at
some length by \citet[][hereafter C09]{C09}, who classified more than
one hundred of them.  The GPs are rare objects, located in
lower-density environments and spectroscopically characterized by very
faint continuum emission and strong optical emission lines, namely
[O{\sc iii}]$\lambda$5007, with an unusually large equivalent width
of up to $\sim$1000\AA.  These properties along with their extreme
compactness (unresolved in Sloan Digital Sky Survey (SDSS) imaging) 
are the reason for their green colors and point-like appearance in 
the SDSS plates. According to C09, GPs are low-mass galaxies (stellar masses
$M_{\star}$$\sim$10$^{8.5}$ to 10$^{10}$$M_{\odot}$) with extremely high
star formation rates (SFRs up to 30 $M_{\odot}$ yr$^{-1}$), as derived
from their emission lines and UV luminosity.  Therefore, their
specific SFRs (SSFRs, i.e., SFR per unit of mass) are among the highest
inferred in the nearby universe.  The GP properties are then
consistent with the picture of galaxy downsizing \citep{Cowie96},
where the sites of active star formation shift from high-mass galaxies
at early times to lower mass systems at later epoch \citep[see
  also][]{Bundy06,Cowie08,Noeske07}.

The motivation for this Letter was one of the most striking results
obtained by C09: the nearly constant solar metallicity of the GPs
(log[O/H]+12$\sim$8.7 in median) over more than two orders of
magnitudes in stellar mass.  These results would imply strong
constrains to the nature of GPs and their evolutionary status.  In
this Letter we examine in more detail the GP chemical abundances,
namely oxygen and nitrogen, and their relation with the stellar mass,
using the same SDSS spectra but a different methodology than in
C09. This analysis indicates instead, that GPs can be considered as a
genuine population of metal-poor galaxies.  The results are compared
with a larger sample of local SFGs from the SDSS and with a smaller 
sample of super compact UV-luminous starburst
galaxies \citep{Heckman05} placed at the same redshift range of the
GPs.  In particular, the position of GPs in the fundamental relation
between mass and metallicity is discussed in the context of their
starburst activity and feedback processes.

\section{THE DATA}

The sample of galaxies was taken from the list of C09.  They compiled
and analyzed by means of SDSS Data Release 7
\citep[SDSS DR7;][]{York} imaging and spectroscopy, {\em Hubble Space 
Telescope (HST)} optical and {\em GALEX} UV imaging, a well-defined 
sample of GPs in the redshift range $0.11 \leq z \leq 0.36$ based on 
color selection criteria.  From
their list we selected a sub-sample of 79 GPs spectroscopically
classified as star-forming systems (their Table~4), ruling out active
galactic nuclei (AGNs) according to the diagnostic diagram
[\nii]/\ha\ versus [\oiii]/\hb\ \citep[][]{BPT}.  We refer the reader to
C09 for a complete description of the sample selection.

We obtained the full set of calibrated one-dimensional spectra, 
i.e., {\sl spSpec fits} files, from the SDSS archive using the SDSS DR7 query
tools\footnote{\url{http://cas.sdss.org/dr7/en/tools/search/sql.asp}
  and \url{http://das.sdss.org}}. The SDSS spectra cover
3800\AA$<\lambda<$9200\AA\ and the fiber has a 3$''$ diameter which, due
to the compactness of the galaxies, include $\ga$70\% of the GP
$r-$band emission\footnote{We used $f_{fib}=10^{-0.4(m_{r,
     \rm{fiber}}-m_{r, \rm{Petro}})}$, which is the fraction of the fiber total
  light (line+continuum) within the fiber aperture}.  Emission-line
integrated fluxes were consistently measured for each galaxy spectrum
using the {\sc IRAF} task {\sl splot}, by summing all the pixels of
the emission line after a linear subtraction of the continuum. The
flux uncertainties are then calculated as

\begin{equation}
\sigma_{l} = \sigma_{c} \sqrt{N + \frac{EW}{\Delta}}
\end{equation}
\citep{PMD03}, where $\sigma_{l}$ is the line flux uncertainty,
$\sigma_{c}$ is the uncertainty in the position of the continuum, $N$
is the number of pixels in the flux measurement, $EW$ is the
equivalent width of the line and $\Delta$ is the wavelength
dispersion.  In most cases the measurements are in agreement with
those based on Gaussian fitting.

\section{RESULTS}

\subsection{Green Peas: A Genuine Population of Metal-poor Galaxies}

Physical properties and oxygen and nitrogen ionic abundances were
calculated from the measured emission lines using the direct method,
which is expected to provide more accurate abundances compared to
strong-line methods \citep[see][hereafter PMC09]{Enrique09}.  First,
we have estimated the reddening constant from the \ha/\hb\ ratio using
the extinction curve of \citet{Cardelli89}.  Then, for those objects
with a reliable measurement of the [\oiii] $\lambda$ 4363 emission
line ($\sim$ 70\% of the sample), we have derived the [\oiii] electron
temperature.  The [\oii] electron temperature has been derived from it
using the relations proposed by PMC09. O$^+$, O$^{2+}$, and N$^+$ ionic
abundances have been calculated using these temperatures and the
relative intensities of the corresponding brightest emission lines.
Finally, total O/H and N/O abundance ratios have been obtained with
the usual expressions (see PMC09).

For the full sample, including those galaxies with no direct
determination of the temperature, oxygen total abundances have 
also been estimated from the strong-line calibrator N2 $\equiv$
\mbox{log([\nii] $\lambda$6584/\ha)}.  We have used the empirical
relation derived by PMC09, which gives oxygen abundances fully
consistent with those derived from $T_{\rm e}$-sensitive methods over
the whole observed range of metallicity.  According to these authors,
N2 has a non-negligible dependence on the N/O ratio. Nevertheless, our
calculations indicate that correction for this dependence in this
sample does not reduce the dispersion, so we have preferred to keep
the calculated abundances in the non-corrected form. Regarding N/O,
this can be derived using the N2S2 parameter (i.e., [\nii]/[\sii]),
minimizing its dependence on reddening or flux calibration.

\begin{figure}[t!]
\begin{tabular}{l l}
\includegraphics[scale=.35]{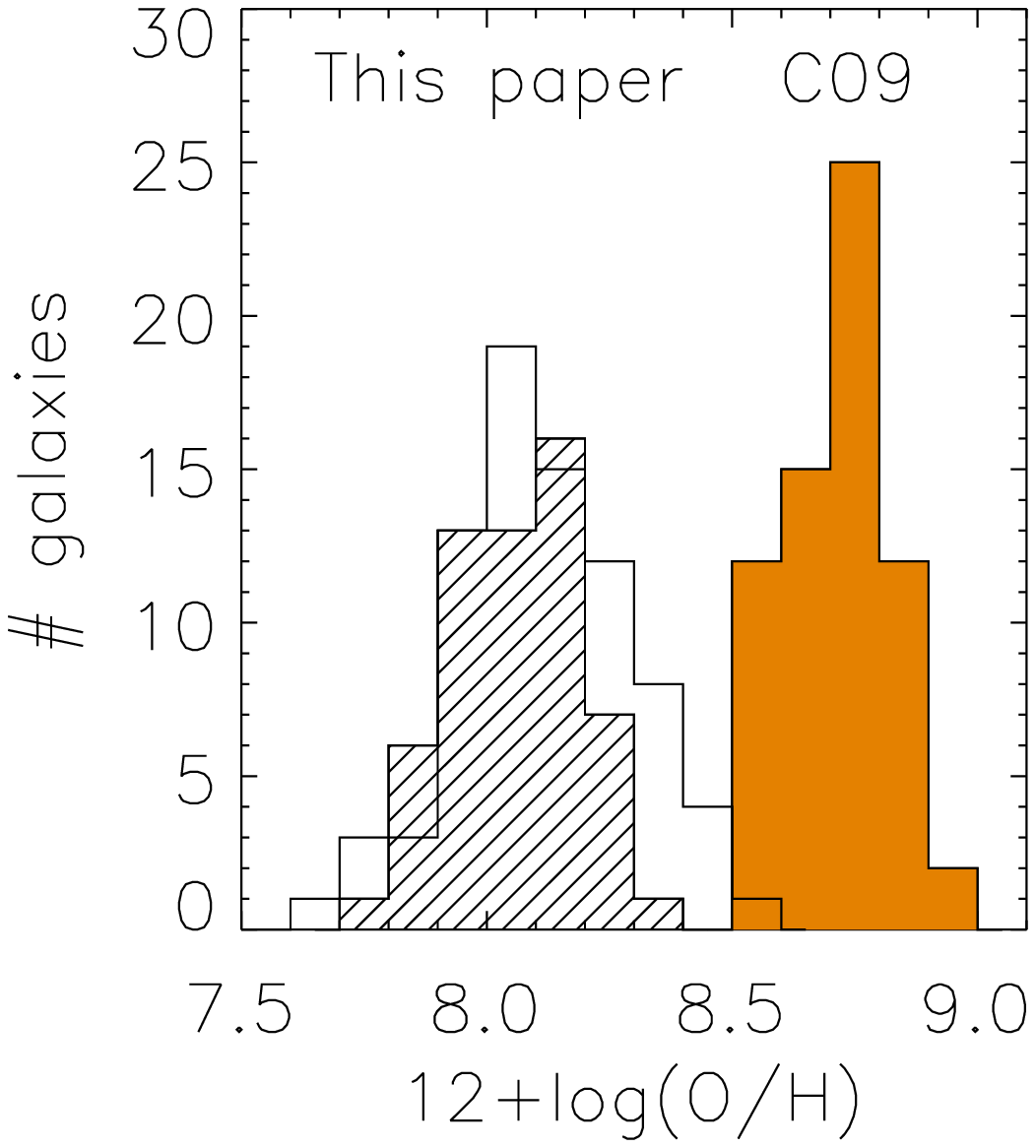}&
\includegraphics[scale=.35]{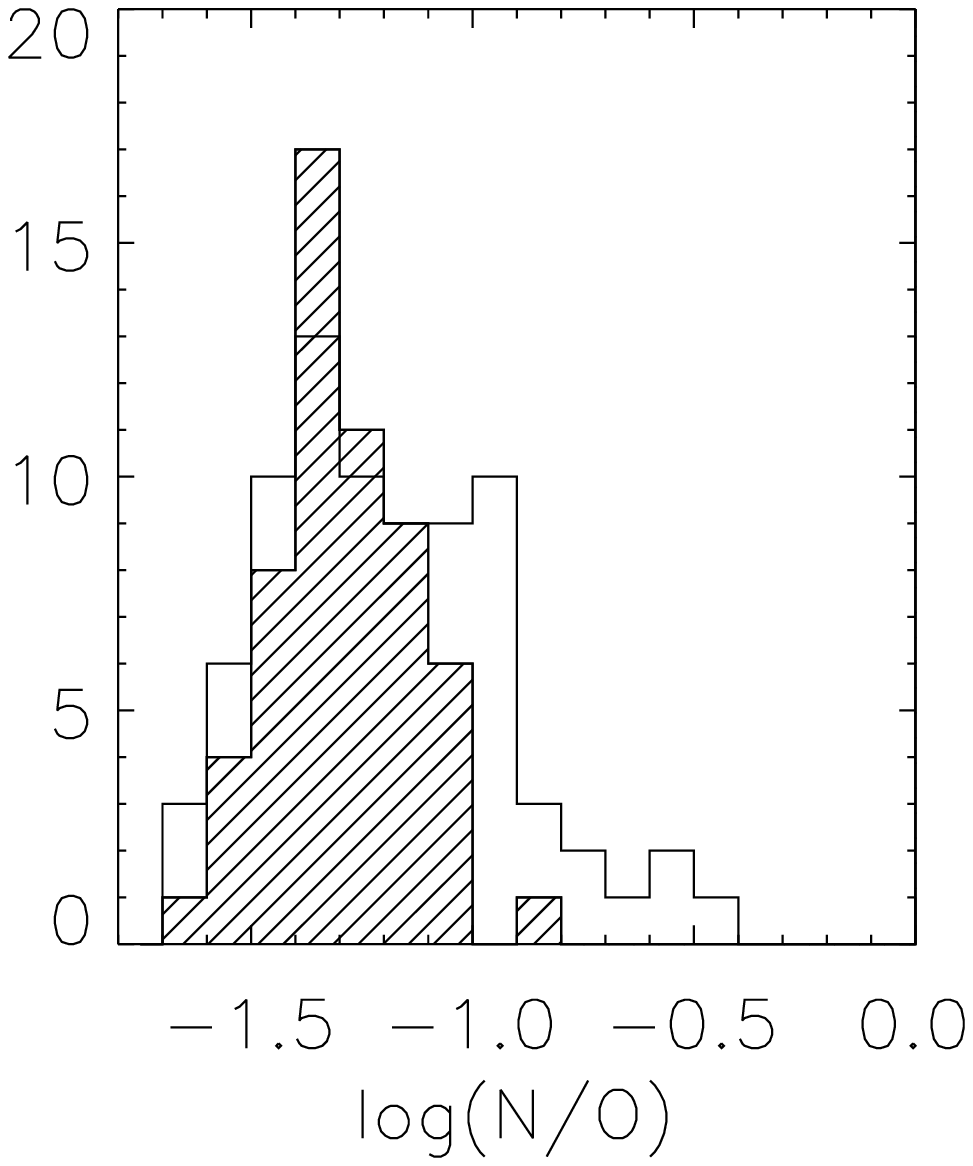}
\end{tabular}
\caption{Oxygen and nitrogen abundances.  On the left, the lined 
and the non-lined histograms indicate the number of GPs with metallicities 
calculated from the direct method and using the N2 index, respectively. 
For comparison, metallicities derived by C09 are shown in the filled 
histogram. On the right, the lined and the non-lined histograms show the
corresponding N/O distributions. (A color version of this figure is available 
in the online journal.)\label{fig1}}
\end{figure}

We have compared the derived properties of the GPs with two different
samples of SFGs.  First, we have taken the large
sample of emission-line galaxies listed in the Max Planck
Institute for Astrophysics/Johns Hopkings University (MPA/JHU) Data 
catalog of the SDSS DR 7\footnote {Available at
  http://www.mpa-garching.mpg.de/SDSS/ }, which covers a redshift
range 0.03 $\leq z \leq$ 0.37.  We have kept the non-duplicated galaxies
with a signal-to-noise ratio (S/N) of at least 5 in all the involved
lines.  Then, we have discarded those objects classified as AGNs,
using the diagnostic diagram [\nii]/\ha\ versus [\oiii]/\hb.  Finally, we
have calculated oxygen and nitrogen abundances in the same way as
described above for the sample of GPs.  Moreover, using the SDSS SFGs
with direct estimation of the ionic abundances, we have obtained the
empirical relation between N/O and the N2S2 parameter (PMC09) that
was consistently applied to estimate N/O for all the galaxies used in
our analysis. Our polynomial fit:
\begin{equation}
\log ({\mathrm N}/{\mathrm O}) = -0.76 + 1.94\, {\mathrm N}2{\mathrm
  S}2 + 0.55\, ({\mathrm N}2{\mathrm S}2)^2
\end{equation}
where N2S2 $=$ log([\nii]/[\sii]), gives a slightly higher slope and a
much lower dispersion ($\sim$0.1 dex) than the calibration of PMC09.

In addition, we have taken for comparison a sample of 31 nearby ($z <
0.3$) UV-luminous starburst galaxies studied by \citep{Overzier08,
  Overzier09a, Overzier09b}.  They showed that these galaxies exhibit
strong similarities in their properties with the Lyman-break galaxies
at high--$z$, and therefore were proposed as good local Lyman-break
analogs (LBAs).  Most massive LBAs display a resolved morphology,
characterized by the presence of a dominant central object
\citep[DCO,][]{Overzier09a}.  In contrast, low-mass LBAs are
metal-poor galaxies showing close similarities (e.g., morphology,
redshift, UV luminosities, size, SSFRs, and environment) to GPs (C09).
In fact, three LBAs are also included in the GP sample.  These three
galaxies are those for which {\em HST} optical imaging was used to analyze
the morphology of the GPs (see Figure~7 in C09).  To be consistent, the
oxygen and nitrogen abundances for the sample of LBAs were
re-calculated using the line ratios measured from SDSS spectra
\citep{Overzier09a}, and the same empirical calibrations described
above.

In Figure~\ref{fig1}, we show histograms with our metallicity estimates
for the GPs and those derived by C09 for comparison.  The GPs show
direct metallicities $7.7 \la$ 12$+$log(O/H)$\la 8.4$, with a mean
value of 8.05 $\pm$ 0.14, i.e., $\sim$20\% the solar value\footnote{We
  adopted $12+$log(O/H)$_{\odot}=8.69$ \citep{Allende01}}, whereas
their measured C$($\hb$)$ and \ha/\hb\ mean values are 0.23 and 3.3,
respectively.  Our direct estimates and those derived using the N2
parameter agree well in $\sim$0.05 dex. While
extinction values are in agreement with those obtained by C09,
Figure~\ref{fig1} shows a clear offset of $\sim$0.65 dex between their
metallicity estimates (mean log(O/H) $+ 12 =$ 8.7) and ours.  The C09
values were obtained using the \NII/\OII\ ratio to estimate log(O/H)
$+$ 12 via the theoretical calibration based on photoionization
models by \citet{KD02}.  Systematical discrepancies are usually found
in the literature between metallicities derived from theoretical and
empirical methods \citep[][and references therein]{KE08}. We have used
the conversion constants provided by \citet{KE08} to scale our
N2-based oxygen abundance estimates to those expected from the
theoretical calibration (their Table 3).  In doing so, we obtained a
mean difference of 0.23 dex, which is not enough to explain the
largest offset in O/H shown in Figure~\ref{fig1}. We conclude that GPs
are genuine metal-poor galaxies with low extinction values.

\begin{figure}[t!]
\includegraphics[scale=.46]{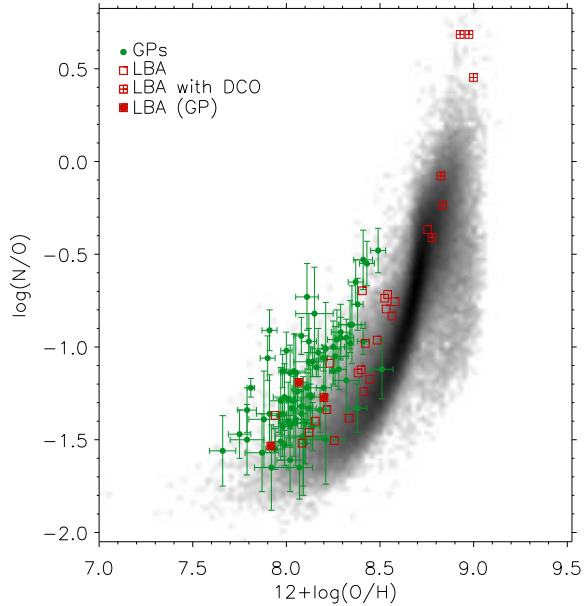}
\caption{N/O vs. O/H. The two-dimensional histogram shows the number 
distribution of SDSS SFGs in the logarithmic gray scale. 
GPs are indicated by green circles, while LBAs are indicated with open 
red squares. LBAs with DCOs are marked with open red squares with plus 
sign, whereas those LBAs also belonging to the GP sample are indicated 
with filled red squares. (A color version of this figure is available 
in the online journal.)\label{fig2}}
\end{figure}

\subsection{N/O Ratios and Metallicity}

The nitrogen-to-oxygen ratio is a powerful evolutionary indicator in
galaxies \citep{Pilyugin04,Molla06}.  We have investigated the
relation between the N/O ratio and the oxygen abundance for the sample
of GPs, LBAs, and the SFGs taken from the SDSS in Figure~\ref{fig2}.
For the SFGs, we found a positive trend with an increasing scatter
toward the metal-rich regime reflecting both primary and secondary
production of nitrogen in the same metallicity range.  At low
metallicities the trend flattens out, possibly a consequence of
nitrogen being of primary origin, coming mainly from massive stars.
In the low-to-intermediate metallicity range, most GPs and some LBAs
display systematically larger N/O ratios for a given metallicity
compared to the SDSS SFGs.  Possible reasons for this include extra
production of primary nitrogen, coming from low-metallicity
intermediate-mass stars \citep{Renzini81,Gavilan06,Molla06}, pollution
by Wolf$-$Rayet stars \citep[e.g.,][]{Brinchmann08,Angel10,AnaMon10}, or
hydrodynamical effects involving outflow and inflow of gas
\citep{Vanzee98,Koppen05}.  This systematically larger N/O ratios
could explain the difference between our metallicity and the higher
estimates obtained by C09 using the calibration by \cite{KD02}.  The
latter assumes that N/O is roughly constant below
12$+$log(O/H)$\sim$8.4, and proportional to O/H above this value.
This implies that larger values of [\nii]/[\oii] lead to higher
metallicities, but this calibration does not take into account the
dependence of [\nii]/[\oii] on the variation of the N/O ratio at a
given O/H. According to PMC09, large values of the N/O ratio can
enhance the [\nii]/[\oii] ratio even in the low metallicity regime.

\subsection{Relation Between Stellar Mass, Metallicity and the N/O Ratio}

\begin{figure*}[t!]
\begin{tabular}{c c}
\includegraphics[scale=.5]{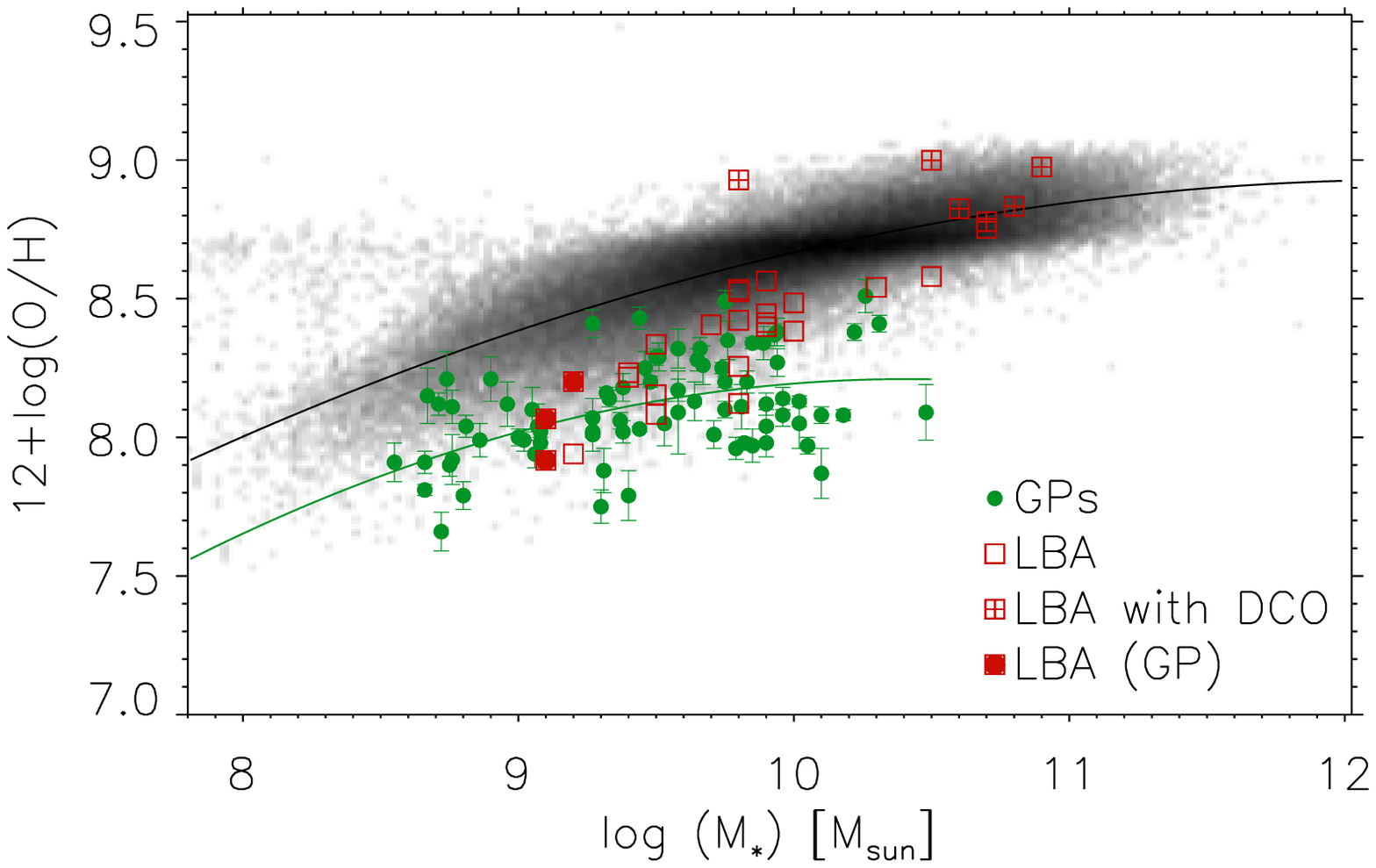}&
\includegraphics[scale=.5]{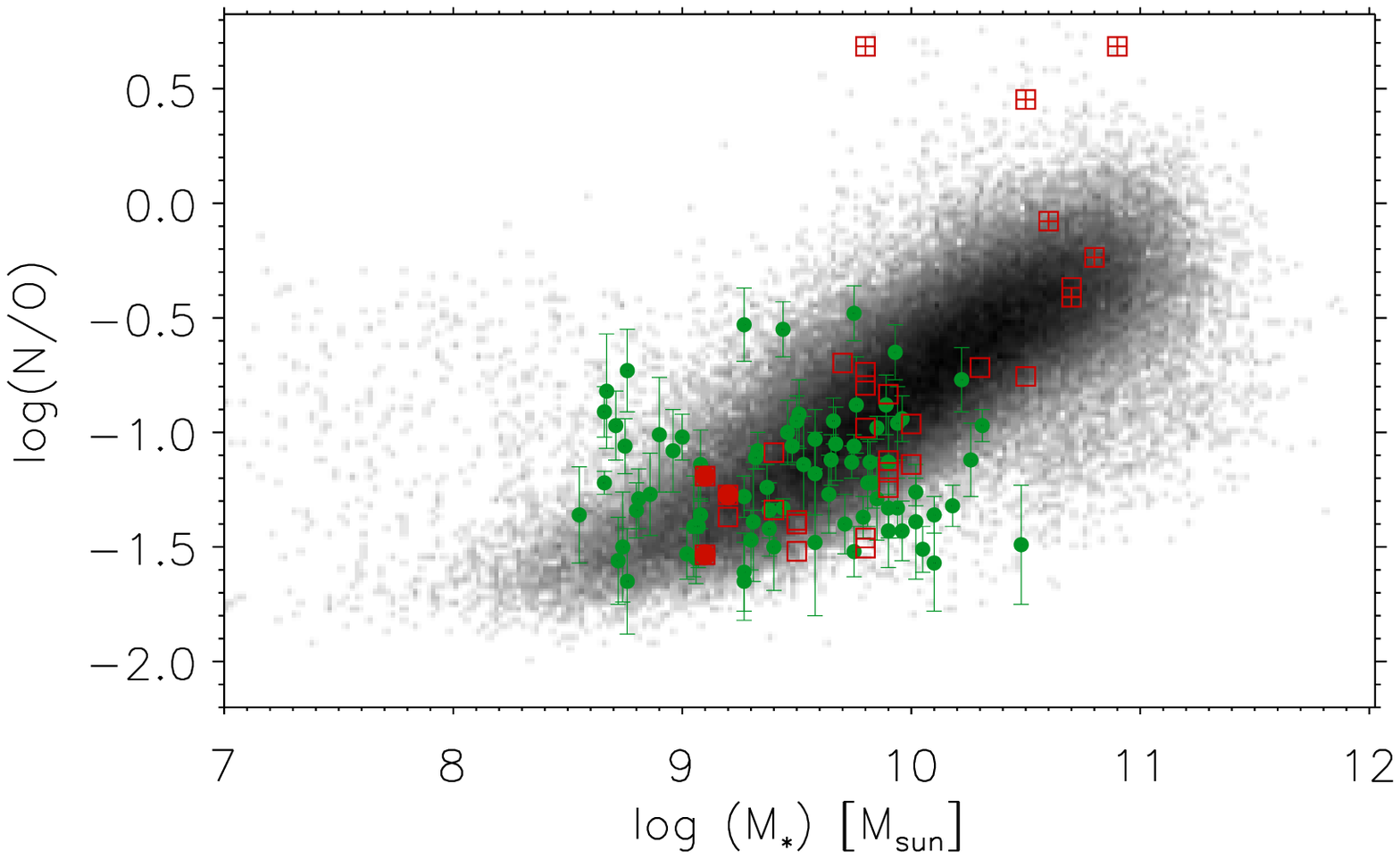}
\end{tabular}
\caption{$M_{\star}$ vs. O/H (left) and $M_{\star}$ vs. N/O 
(rigth). 
Symbols stand for the same type of objects as described in Figure~\ref{fig2}.
For better comparison the trends observed in the $M_{\star}-Z$ plane, green 
and black lines indicating second-order polynomial fits to the 
GPs and SFGs, respectively, have been overplotted.
 (A color version of this figure is available in the online journal.)
\label{fig3}}
\end{figure*}

In Figure~\ref{fig3}, we studied the stellar mass$-$metallicity relation
\citep[$M_{\star}-Z$, MZR;][]{Lequeux79} and the relation between N/O
and the stellar mass ($M_{\star}-$N/O; PMC09) for the GPs, LBAs, and
SDSS SFGs.  Estimates of $M_{\star}$ for the three galaxy samples have
been taken from C09 (GPs), \citet[][LBAs]{Overzier09a}, and the MPA$-$JHU
catalog (SFGs). They were derived using spectral energy distribution 
fitting and SDSS photometry. 
Thus, stellar masses are expected to to be consistent within a typical 
dispersion of $\sim$ 0.20 dex \citep[e.g.,][]{Drory,vanderwel06}. 
This value, is lower than the expected uncertainties of 0.3 dex quoted 
for GPs and LBAs as due to uncertainties in their star formation history
\citep[C09;][]{Overzier09a}.

In the $M_{\star}-$Z plane, SDSS SFGs show a clear positive trend which
flattens toward higher masses and show a relatively large scatter at
lower masses, in agreement with previous findings for local galaxies
\citep[e.g.,][]{Tremonti}.  Similarly, GPs and LBAs increase their
metallicity with increasing masses.  Nevertheless, GPs lie more than a
factor of 2 ($\ga$ 0.3 dex) below the MZR of SDSS SFGs, i.e. {\it at
  a fixed stellar mass GPs are systematically more metal-poor than
  normal SFGs}. A similar shift in the MZR has been observed for the
less massive (log(M$_{\star}) \la 10$) LBAs \citep[][this
  work]{Hoopes07,Overzier09b}.  In terms of stellar mass, the observed
offset (as large as 1 dex) largely exceeds the typical uncertainty
quoted for the GP masses.

On the other hand, the SDSS SFGs display a correlation between N/O and
$M_{\star}$, with higher N/O ratios at higher stellar masses, in
agreement with PMC09 for a different sample of local SFGs.  The
existence of the $M_{\star}-$N/O relation reflects the fact that the most
massive galaxies evolve more quickly and, hence, they should have on
average higher metallicities and N/O ratios. Though the scatter in the
$M_{\star}-$N/O relation is large, most GPs and low-mass LBAs are
roughly consistent with the trend of SDSS SFGs, and no systematic
offset is observed.

\section{DISCUSSION}

The shape of the MZR in galaxies has been found to depend on several
key processes in galaxy evolution, including the efficiency of star
formation \citep[e.g.,][]{Lequeux79,MollaDiaz05,Brooks07,Calura09},
the action of metal-rich selective outflows and metal-poor gas inflows
\citep[e.g.,][]{Larson74,Garnett02,Tremonti,Finlator08}, and the
possible variations in the initial mass function \citep{Koppen07}.
Moreover, different amounts of dark matter can also assist in some of
these mechanisms \citep{Dekel86}.

The GPs follow a relation between mass and metallicity that parallels
the MZR defined by the SDSS SFGs, but is offset $\ga$0.3~dex to lower
metallicities.  Interestingly, we find some remarkable similarities
between the GPs and the population of [\oiii]-selected galaxies of
similar luminosity in the $z$ range 0.29--0.42 recently found by
\citet{Salzer09}.  These galaxies follow a luminosity$-$metallicity
relation that parallels the one defined by SFGs, but is offset by a
factor of more than 10 to lower abundances.  On the other hand, since
GPs and low-mass LBAs lie in a similar offset position in the MZR
relative to normal SFGs, their mentioned similarities appear even
greater. One possibility to explain their offset position in the MZR
is that these galaxies could be still converting a large amount of
their cold gas reservoirs into stars.  In that case, their low
abundances could be due to their relatively young ages compared to
normal SFGs. In the same range of masses, both GPs and LBAs have much
higher SSFR (typically $>$10$^{-9}$yr$^{-1}$) compared to other SFGs
of similar mass (C09).  Recent studies show that galaxies with higher
SSFRs or larger half-light radii for their stellar mass have
systematically lower metallicities \citep[e.g.,][]{Tremonti,
  Ellison08}.  However, we have found even greater under-abundances in
the GPs, which have high SSFRs but are extremely compact.

Some models show that in highly concentrated (typical sizes $<$3~kpc)
low-mass galaxies such as the GPs, galactic winds induced by their
large SSFR are strong enough to escape from their weak potential
wells, diminishing the observed global abundances
\citep[e.g.,][]{Finlator08}. In contrast, analytical models by
\citet{Dalcanton07} show that any subsequent star formation to the
outflow will remove their effects on metallicity, unless galaxies have
an inefficient star formation. Smoothed particle hydrodynamics (SPH) 
plus $N$-body simulations have
shown that low star formation efficiencies, regulated by supernova
(SN) feedback, could be primarily responsible for the lower
metallicities of low-mass galaxies and their overall trend in the MZR
\citep{Brooks07}.  As shown by \citet{Erb06} for SFGs at $z \sim 2$,
the constancy in the offset of the MZR suggests the presence of
selective metal-rich gas loss driven by SN winds.

Inflow of metal-poor gas, either from the outskirts of the galaxy or
beyond, can dilute metals in the galaxy centers, explaining an offset
to lower abundances in both the MZR
\citep{Mihos96,Barnes96,Finlator08} and the N/O -- O/H diagram. In
starburst galaxies, a recent cold gas accretion can be due to
interactions, which eventually increases the gas surface density and
consequently the star formation.  As explained by \citet{Ellison08},
the dilution of metals due to an inflow can be restored by the effects
of star formation depending on the dilution-to-dynamical timescale
ratio.  Since this ratio depends inversely with galaxy radius,
galaxies with smaller radius, such as the GPs, may be expected to take
longer time to enhance their oxygen abundances to the values expected
from the MZR.  In this line, the position of GPs in the
$M_{\star}-$N/O relation and the offset observed in the N/O -- O/H
plane may favor this scenario.  Models by \citet{Koppen05} have shown
that the rapid decrease of the oxygen abundance during an episode of
massive and rapid accretion of metal-poor gas is followed by a slower
evolution which leads to the closed-box relation, thus forming a loop
in the N/O -- O/H diagram.

The inflow hypothesis is also strongly suggested by the disturbed
morphologies and close companions observed in spatially resolved {\em HST}
images for three GPs and most LBAs \citep[C09;][]{Overzier08}.  Recent
results revealed that galaxies involved in galaxy interactions fall
$\ga$0.2 dex below the MZR of normal galaxies due to tidally induced
large-scale gas inflow to the galaxies' central regions
\citep[e.g.,][]{Kewley06,Michel-Dansac08,Peeples09}.  Several
$N$-body/SPH simulations have shown that major interactions drive
starbursts and gas inflow from the outskirts of the H {\sc i}
progenitor disks \citep[e.g.,][]{Mihos96,Rupke10}, also supporting
this scenario.

We conclude arguing that recent interaction-induced inflow of gas,
possibly coupled with a selective metal-rich gas loss driven by SN
winds, may explain our findings and the known galaxy properties.
Nevertheless, further work is needed to constrain this possible
scenario.  In particular, future assessment of the H {\sc i} gas
properties and the star formation efficiency of GPs, as well as the
behavior of effective yields with mass compared with models of
chemical evolution, will shed new light on the relative importance of
the above processes.

Our results allow us to further constrain the evolutionary status of the
GPs. These galaxies, as well as the low-mass LBAs and the
[\oiii]-selected galaxies by \citet{Salzer09} should be analyzed and
compared in more detail to elucidate whether these rare objects are
sharing similar evolutionary pathways.  Even if this is not the case,
their properties suggest that these galaxies are snapshots of an
extreme and short phase in their evolution.  They therefore offer the
opportunity of studying in great detail the physical processes
involved in the triggering and evolution of massive star formation and
the chemical enrichment processes, under physical conditions
approaching those in galaxies at higher redshifts.

Forthcoming analysis, based on high S/N intermediate/high-resolution
spectroscopy and deep NIR imaging with the Gran Telescopio Canarias (GTC), 
will be used to better constrain the evolutionary status of the GPs.

\acknowledgments

We are very grateful to P. Papaderos, M. Moll\'a and Y. Tsamis for
very stimulating discussions and useful suggestions to improve this
manuscript. We also thank the anonymous referee for a useful and prompt 
report. This work has been funded by grants AYA2007-67965-C03-02,
and CSD2006-00070: First Science with the GTC ({\url
  http://www.iac.es/consolider-ingenio-gtc/}) of the
Consolider-Ingenio 2010 Program, by the Spanish MICINN.

{\it Facility:} \facility{Sloan}

\end{document}